\documentclass[twocolumn,conference]{IEEEtran}

\IEEEoverridecommandlockouts

\usepackage{color}
\usepackage{graphicx}
\usepackage{amsmath}
\usepackage{amssymb}
\usepackage{algorithm}
\usepackage{algorithmic}
\usepackage{amsmath}
\usepackage{multirow}
\usepackage{booktabs}
\usepackage{array}
\usepackage{amsthm}
\usepackage{stfloats}
\usepackage{caption}
\usepackage{bm}
\usepackage{epstopdf}
\usepackage{setspace}
\usepackage{flushend}

\newcommand{\be}{\begin{equation}}
\newcommand{\ee}{\end{equation}}
\newcommand{\bea}{\begin{eqnarray}}
\newcommand{\eea}{\end{eqnarray}}
\newcommand{\ba}{\begin{array}}
\newcommand{\ea}{\end{array}}

\makeatletter

\newcommand{\Rmnum}[1]{\expandafter\@slowromancap\romannumeral #1@}
\makeatother

\newcommand{\RNum}[1]{\uppercase\expandafter{\romannumeral #1\relax}}

\title{A Practical Beamforming Design for Active RIS-assisted MU-MISO Systems
\thanks{$*$  Corresponding author.} \thanks{
This work is supported in part by the National Natural Science Foundation of China (Grant No. 62371090 and 62071083), in part by Liaoning Applied Basic Research Program (Grant No. 2023JH2/101300201), and in part by Dalian Science and Technology Innovation Project (Grant No. 2022JJ12GX014).}
}
\author{\IEEEauthorblockN{Yun Yang$^{\dag}$, Zhiping Lu$^{\ddag}$, Ming Li$^{\dag \ast}$, Rang Liu$^{\S}$, and Qian Liu$^{\dag}$
\vspace{-0.0 cm} }

\IEEEauthorblockA{$^{\dag}$Dalian University of Technology, Dalian, Liaoning 116024, China}
\IEEEauthorblockA{$^{\ddag}$Beijing University of Posts and Telecommunications, Beijing 100876, China \\
State Key Laboratory of Wireless Mobile Communications(CICT), Beijing 100191, China}
\IEEEauthorblockA{$^{\S}$ University of California, Irvine, CA 92697, USA \\ E-mail: \texttt{yunyang@mail.dlut.edu.cn}, \texttt{luzhp$\_$007@163.com}, \texttt{rangl2@uci.edu},\\
\texttt{\{mli, qianliu\}@dlut.edu.cn} }}
\begin{document}
\maketitle
\thispagestyle{empty}
\begin{abstract}
Reconfigurable Intelligent Surfaces (RIS) have been proposed as a revolutionary technology with the potential to address several critical requirements of 6G communication systems. Despite its powerful ability for radio environment reconfiguration, the ``double fading'' effect constricts the practical system performance enhancements due to the significant path loss. A new active RIS architecture has been recently proposed to overcome this challenge. However, existing active RIS studies rely on an ideal amplification model without considering the practical hardware limitation of amplifiers, which may cause performance degradation using such inaccurate active RIS modeling. Motivated by this fact, in this paper we first investigate the amplification principle of typical active RIS and propose a more accurate amplification model based on amplifier hardware characteristics. Then, based on the new amplification model, we propose a novel joint transmit beamforming and RIS reflection beamforming design considering the incident signal power on practical active RIS for multiuser multi-input single-output (MU-MISO) communication system. Fractional programming (FP), majorization minimization (MM) and block coordinate descent (BCD) methods are used to solve for the complex problem. Simulation results indicate the importance of the consideration of practical amplifier hardware characteristics in the joint beamforming designs and demonstrate the effectiveness of the proposed algorithm compared to other benchmarks.
\end{abstract}

\begin{IEEEkeywords}
Active RIS, practical joint beamforming design, incident power.
\end{IEEEkeywords}

\maketitle
\section{Introduction}
In recent years, with the rapid development of advanced technologies, the sixth-generation (6G) communication is expected to achieve better ubiquitous wireless communications by reconfiguring the radio propagation environment. In this light, reconfigurable intelligent surface (RIS) has drawn significant interest as a revolutionary technology to help fulfill various requirements of 6G communication systems.

Specifically, RIS is a planar surface consisting of low-cost passive reflective elements. Each element is capable of independently inducing controllable amplitude and/or phase changes to the incident signal, which results in modifying the wireless channel and enabling an intelligent and programmable wireless environment \cite{3towards}. However, a fatal ``double fading" effect occurs, which is inherently a multiplicative fading in cascading channel composed of base station (BS)-RIS and RIS-user links.
This results in significant path loss from reflected RIS signals being orders of magnitude, greater than the unobstructed direct link. Consequently, only marginal performance gains can be achieved when RIS is not placed close enough to users or BS in practical deployment.

To effectively tackle this ``double fading" effect issue, the authors in \cite{5activeRIS} proposed active RIS architecture that equips each element with a reflection amplifier, which can amplify the signal reflected by the RIS at the electromagnetic (EM) level. Compared with traditional passive RIS, active RIS not only adjusts the phase-shift of the incident signal but also amplifies the weak incident signal after propagating via the BS-RIS channel, which overcomes the impact of the ``double fading'' effect. Without the need for an expensive and high power consumption radio frequency (RF) chain, active RIS can be implemented by hardware-efficient active components \cite{4GHZ},
such as tunnel diodes \cite{8tunnel}.

Based on the aforementioned advantages, active RIS has been intensively investigated in recent literature \cite{9prevail}-\cite{15SWIPT}. In \cite{9prevail}, the authors developed a signal model for active RIS and demonstrated that active RIS can achieve a prominent sum-rate gain of 130\%, while passive RIS only achieves 22\% enhancement compared with traditional wireless system without RIS. In \cite{10wireless}, the authors optimized active or passive RIS deployment by taking downlink/uplink communication into account in order to maximize the transmission rate. In \cite{11sub}, the authors proposed a sub-connected architecture to reduce the high hardware cost and power consumption caused by the fully-connected architecture. A realistic active RIS power consumption model was employed to improve the energy efficiency of active RIS-assisted systems. Secure transmission, unmanned aerial vehicles (UAV) communication and simultaneous wireless information and power transfer (SWIPT) communication have also been investigated in \cite{13secure}-\cite{15SWIPT}, respectively.
\begin{figure}[t]
\centering
  \includegraphics[width=0.42\textwidth]{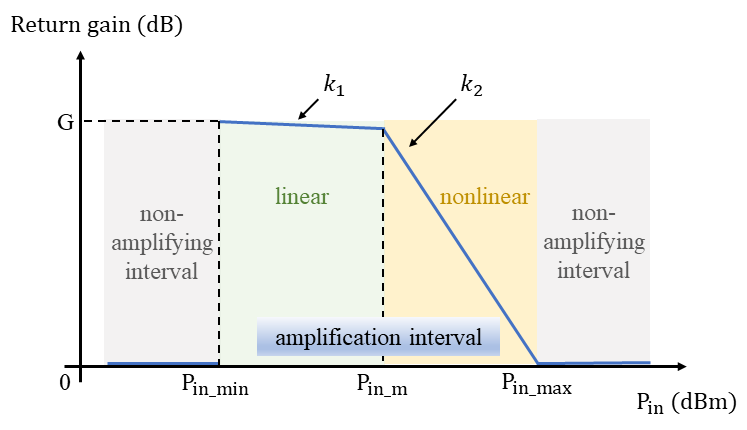}
  \caption{Return gain of reflection amplifier versus the incident signal power.}
  \label{fig:interval}
  \vspace{-0.5cm}
\end{figure}

All of the above studies are based on an ideal reflective amplification signal model, which has a constant amplification factor. Thus, the existing active RIS beamforming designs do not consider the actual response of the amplification circuit with different strength levels of incident signals. However, due to the hardware characteristics and limitations of the reflection amplification circuit, the amplifier can efficiently amplify the incident signal without distortion only when the power of the incident signal falls within a so-called linear amplification interval. More specifically, as the incident signal power is within such an interval, the reflection gain is constant or varies slightly with the incident signal power. When the incident signal power falls out of this specific interval, due to the characteristics of the amplifier circuit, the device has nonlinear amplification with significant distortion, or even only reflects signal without amplification. Unfortunately, existing active RIS studies ignore these important amplifier hardware characteristics, that exert severe inaccuracy on the amplification properties of active RIS and cause performance degradation when using inaccurate modeling in the active RIS optimizations.

In order to facilitate the deployment of active RIS in practical wireless systems, in this paper we first investigate the amplification principle of typical active RIS and introduce a more accurate and practical amplification model based on the amplifier characteristics. Then, based on the new amplification model, we propose a joint transmit beamforming and RIS reflection beamforming design algorithm for an active RIS-assisted multiuser multi-input single-output (MU-MISO) communication system. Specifically, we aim to maximize the sum-rate of the active RIS-assisted MU-MISO system by jointly designing the transmit beamforming vectors at the BS, the phase-shift matrix and updating the amplification factor matrix according to the incident signal power at the active RIS, while satisfying the transmit power constraint of the BS, the power consumption constraint on each active RIS element, and the power constraints of the incident signals on the active RIS. In an effort to solve this complicated problem, fractional programming (FP), majorization minimization (MM) and block coordinate descent (BCD) methods are leveraged to optimize the variables alternatively. The simulation results illustrate the importance of the utilization of the introduced active RIS model by considering practical amplifier characteristics. Moreover, the effectiveness of the proposed practical joint beamforming design algorithm is also verified in boosting the sum-rate of the active RIS-assisted MU-MISO communication system.

\section{Amplification Principle and System Model}

In this section, we will first elaborate on the amplification principle of the reflection amplifier in active RIS. Then, with the aid of the amplification principle of active RIS, we present a practical signal and transmission model for an active RIS-assisted MU-MISO communication system.

\subsection{Amplification Principle} \label{Amplify}
A tunnel diode is being used to fabricate reflection amplifiers with high gain on account of its low bias requirements and negative differential resistance properties. The tunnel diode-based reflection amplifier is designed on the concept of an oscillator, which will lock on to the frequency of the input signal when appropriately biased. It is implemented on a microstrip line. When the impedance of the reflection amplifier is matched to that of the microstrip line, the reflection amplifier converts the DC power supplied by the bias source into RF power in the form of \textit{injection locking}, which is augmented to the incident signal as it is reflected. This is in accordance with the principle of conservation of energy \cite{8tunnel}, \cite{hardware}.

However, both the occurrence of locking and the impedance of the reflection amplifier are related to the amplitude of the incident signal. Therefore, as illustrated in Fig. \ref{fig:interval}, for the incident signal power located in various intervals, the reflection amplifier will exhibit the following characteristics \cite{8tunnel}:

\begin{itemize}
  \item
  \textbf{Linear amplification characteristic:}

  As $P_\text{in} \in \left[ P_{\mathrm{in\_min}}, P_{\mathrm{in\_m}} \right] $, the tunnel diode operates in the negative differential resistance region and the reflection amplifier allows linear amplification of the incident signal. Therefore, when the incident signal power varies with this range, the return gain remains consistent or slightly changed. Namely, $k_{1}$ in Fig. \ref{fig:interval} has a 0 or tiny value.

  \item
  \textbf{Nonlinear amplification characteristic:}

  As $P_\text{in} \in \left[ P_{\mathrm{in\_m}}, P_{\mathrm{in\_max}} \right] $, due to the shift of the tunnel diode operating point from the negative differential resistance region to the positive differential resistance region, the power of the output signal through the reflection amplifier no longer varies with the incident signal. Namely, the amplifier exhibits nonlinear amplification characteristics and the return gain has a negative correlation with the incident signal power, i.e. $k_{2} \textless 0$ in Fig. \ref{fig:interval}.

  \item
  \textbf{Reflective and non-amplifying characteristic:}

   As $P_\text{in} \notin \left[ P_{\mathrm{in\_min}}, P_{\mathrm{in\_max}} \right] $, owing to the mismatch between the impedance of the tunnel diode and the microstrip line, the reflection amplifier can only reflect the signal without amplification. In addition, the incident signal power is also too weak to trigger the \textit{lock-in} phenomenon, consequently the reflection amplifier cannot perform the amplification function properly either.
\end{itemize}



\begin{figure}[!t]
\centering
  \includegraphics[width=0.40\textwidth]{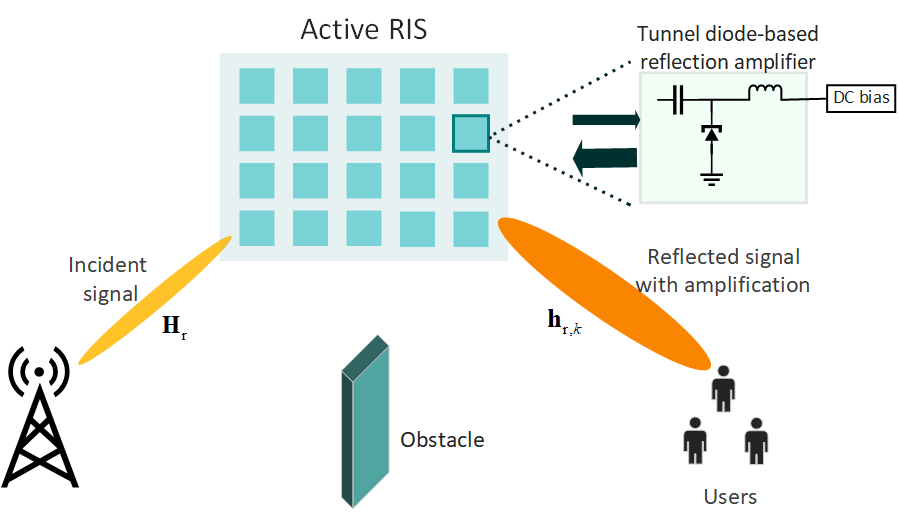}
  \caption{The active RIS-assisted MU-MISO system.}
  \label{fig:System model}
  \vspace{-0.2cm}
\end{figure}

\subsection{System Model}
As shown in Fig. \ref{fig:System model}, we consider an active RIS-assisted downlink MU-MISO communication system, where an $M$-antenna BS serves $K$ single-antenna users. Specifically, the active RIS consists of $L$ elements, each of which is integrated with a reflection amplifier. We suppose that the direct channel from the BS to users is blocked and the virtual line-of-sight (LoS) channel is established with the aid of active RIS. Thus, the incident signal on the $l$-th element of active RIS is expressed as
\begin{equation}
x_{l} = \mathbf{h}_{\mathrm{r}, l}^{H}\sum_{i=1}^{K}\mathbf{w}_{i} s_{i},
\end{equation}
where $\mathbf{h}_{\mathrm{r}, l} \in \mathbb{C}^{M}$ denotes the channel from the BS to the $l$-th element, $\mathbf{w}_{i} \in \mathbb{C}^{M}$ denotes the BS beamforming vector for the symbol $s_{i}$ of the $i$-th user, and $\mathbb{E}\{|s_{i}|^{2}\}=1$. The incident signal on the entire active RIS is expressed as
\begin{equation}
\mathbf{x} = \mathbf{H}_{\mathrm{r}}\sum_{i=1}^{K}\mathbf{w}_{i} s_{i},
\end{equation}
where $\mathbf{H}_{\mathrm{r}}\triangleq[\mathbf{h}_{\mathrm{r}, 1}~\mathbf{h}_{\mathrm{r}, 2}~\ldots\mathbf{h}_{\mathrm{r}, L}]^H \in \mathbb{C}^{L \times M}$ denotes the channel from the BS to the active RIS. Then, the output signal $\mathbf{y}_{\mathrm{ar}}$ amplified and reflected by active RIS is written as
\begin{equation}
    \mathbf{y}_{\mathrm{ar}}=\mathbf{A} \boldsymbol{\Phi}(\mathbf{x}+\mathbf{v}),
\end{equation}
where
$\mathbf{A} \triangleq \operatorname{diag}(\mathbf{a})$,
$\mathbf{a} \triangleq[\sqrt{a_{1}}, \sqrt{a_{2}}, \ldots, \sqrt{a_{L}}]^{T}$
denotes the amplification factor vector of the active RIS,
$\boldsymbol{\Phi} \triangleq \operatorname{diag}(\mathbf{\phi})$,
$\mathbf{\phi} \triangleq[\phi_{1}, \phi_{2}, \ldots, \phi_{L}]^{T} \triangleq[e^{j \theta_{1}}, e^{j \theta_{2}}, \cdots, e^{j \theta_{L}}]^{T}$
denotes the reflection phase-shift vector, and $\mathbf{v} \in \mathbb{C}^{L}$ represents the introduced and amplified noise by reflection amplifier with $\mathbf{v} \sim \mathcal{C} \mathcal{N}\left(\mathbf{0}, \sigma_{v}^{2} \mathbf{I}_{L}\right)$. Therefore, the received signal at the $k$-th user, $k=1, \ldots, K$ can be modelled as
\begin{equation}
\begin{aligned}
    y_{k} & =\mathbf{h}_{\mathrm{r}, k}^{H} \mathbf{y}_{\mathrm{ar}}+n_{k} \\
          & =\mathbf{h}_{\mathrm{r}, k}^{H} \mathbf{A} \boldsymbol{\Phi} \mathbf{h}_{\mathrm{d,r}} \sum_{i=1}^{K} \mathbf{w}_{i} s_{i}+\mathbf{h}_{\mathrm{r}, k}^{H} \mathbf{A} \boldsymbol{\Phi} \mathbf{v}+n_{k},
\end{aligned}
\end{equation}
where $\mathbf{h}_{\mathrm{r}, k} \in \mathbb{C}^{L}$ denotes the channel from the active RIS to user-$k$ and $n_{k}$ represents the additive white Gaussian noise (AWGN) at the $k$-th user with zero mean and variance $\sigma^{2}$.

The signal-to-interference-plus-noise ratio (SINR) of the $k$-th user can be formulated as
\begin{equation}
    \gamma_{k}=\frac{|\overline{\mathbf{h}}_{k}^{H} \mathbf{w}_{k}|^{2}}{\sum_{i \neq k}^{K}|\overline{\mathbf{h}}_{k}^{H} \mathbf{w}_{i}|^{2}+\|\mathbf{h}_{\mathrm{r}, k}^{H} \mathbf{A} \boldsymbol{\Phi}\|^{2}\sigma_{v}^{2}+\sigma^{2}},
\end{equation}
where
$\overline{\mathbf{h}}_{k}^{H} \triangleq \mathbf{h}_{\mathrm{r}, k}^{H} \mathbf{A} \boldsymbol{\Phi} \mathbf{H}_{\mathrm{r}}$
 denotes the equivalent channel from the BS to the $k$-th user.

In this paper, we consider a practical active RIS scheme that each element of the active RIS manipulates the incident signal using its own reflection amplifier circuit. Accordingly, the incident signal power on the $l$-th element is expressed as
\begin{equation}
\label{pin,l}
    p_{\mathrm{in},l} = \sum_{i=1}^{K}\left|\mathbf{h}_{\mathrm{r}, l}^{H}\mathbf{w}_{i}\right|^{2}.
\end{equation}
As discussed in Section \ref{Amplify}, $p_{\mathrm{in},l}$ is required to be in an interval that triggers the amplification mechanism of the reflective amplifier, allowing the active RIS to behave efficiently and reveal its merits. Due to this restriction, we attach the following constraint to guarantee that the incident signal power is amplified without distortion as
\begin{equation}
    P_{\mathrm{in\_min}} \leq \sum_{k=1}^{K}\big|\mathbf{h}_{\mathrm{r}, l}^{H} \mathbf{w}_{k}\big|^{2} \leq P_{\mathrm{in\_m}}, \quad \forall l.
\end{equation}
Then the output power of the $l$-th element is given by
\begin{equation}
    p_{\mathrm{out},l} = \left|\sqrt{a_{l}}\phi_{l}\right|^{2}\left(p_{\mathrm{in},l}+\sigma_{v}^{2}\right).
\end{equation}


\section{Practical Joint Beamforming Design for Sum-rate Maximization Problem}

\subsection{Problem Formulation}
In this section, we propose to maximize the sum-rate of the active RIS-assisted MU-MISO downlink communication system by jointly designing the transmit beamforming vectors $\mathbf{w}_k$ at the BS, the phase-shift matrix $\mathbf{\Phi}$ and updating the amplification factor matrix $\mathbf{A}$ at the active RIS, while guaranteeing the transmit power constraint of the BS, the power consumption constraint on each active RIS element, and the power constraints of the incident signals on the active RIS. The optimization problem can thus be formulated as follows:
\begin{subequations}
\label{ori}
\begin{align}\label{ori a}
    \max _{\mathbf{w}_{k}, \boldsymbol{\Phi}, \mathbf{A}}~&~ \sum_{k=1}^{K} \log _{2}\left(1+\gamma_{k}\right)\\
    \label{ori b}
    \text { s.t. } & ~\sum_{k=1}^{K}\left\|\mathbf{w}_{k}\right\|^{2} \leq P_{\mathrm{BS}}, \\
    \label{ori c}
    & ~\left|\sqrt{a_{l}} \phi_{l}\right|^{2}\Big(\sum_{k=1}^{K}|\mathbf{h}_{\mathrm{r}, l}^{H} \mathbf{w}_{k}|^{2}+\sigma_{v}^{2}\Big) \leq P_{l}, ~~ \forall l, \\
    \label{ori d}
    & ~P_{\mathrm{in\_min}} \leq \sum_{k=1}^{K}|\mathbf{h}_{\mathrm{r}, l}^{H} \mathbf{w}_{k}|^{2} \leq P_{\mathrm{in\_m}}, ~~ \forall l,
\end{align}
\end{subequations}
where $P_{\mathrm{BS}}$ denotes the transmit power budget at the BS, $P_{l}$ denotes the power budget of the $l$-th element on the active RIS, and (\ref{ori d}) guarantees that the signal power incident on the $l$-th element is in the locking interval of the reflection amplifier.

We notice that problem (\ref{ori}) is a challenging non-convex problem since the objective (\ref{ori a}) and the constraint (\ref{ori d}) are non-convex with variables that are highly coupled. In order to effectively solve this problem, we propose to recast the original problem using the FP algorithm and utilize the MM as well as BCD methods to optimize the variables alternatively.

\subsection{Proposed Practical Joint Beamforming Design}
It can be observed that $\mathbf{A}$ and $\mathbf{\Phi}$ always appear as a product form. For simplicity, we define the amplified reflection matrix of active RIS as
\begin{equation}
   \boldsymbol{\Psi} \triangleq \mathbf{A} \boldsymbol{\Phi}=\operatorname{diag}\left(\sqrt{a_{1}} \phi_{1}, \ldots, \sqrt{a_{L}} \phi_{L}\right),
\end{equation}
and define $\boldsymbol{\psi} \triangleq[\sqrt{a_{1}} \phi_{1}, \sqrt{a_{2}} \phi_{2}, \ldots, \sqrt{a_{L}} \phi_{L}]^{T}$, i.e., $\boldsymbol{\Psi}=\operatorname{diag}(\boldsymbol{\psi})$. By applying the \textit{Lagrangian Dual Transform} \cite{FP}, complex fractional terms can be separated from the $\log (\cdot)$ function. The objective in (\ref{ori}) is equivalently transformed to
\begin{equation}
\label{f1}
\begin{aligned}
& \hspace {-2pc} f_{1}\left(\mathbf{w}_{k}, \boldsymbol{\Psi}, \boldsymbol{\rho}\right) = \sum_{k=1}^{K} \log _{2}\left(1+\rho_{k}\right)-\sum_{n=1}^{K} \rho_{k} \quad \\
& +\sum_{k=1}^{K} \frac{\left(1+\rho_{k}\right)\big|\overline{\mathbf{h}}_{k}^{H} \mathbf{w}_{k}\big|^{2}}{\sum_{i=1}^{K}\big|\overline{\mathbf{h}}_{k}^{H} \mathbf{w}_{i}\big|^{2}+\big\|\mathbf{h}_{\mathrm{r}, k}^{H} \boldsymbol{\Psi}\big\|^{2} \sigma_{v}^{2}+\sigma^{2}} ,
\end{aligned}
\end{equation}
where $\boldsymbol{\rho}\triangleq[\rho_{1}, \rho_2, \cdots, \rho_{K}]^{T}$ is introduced as an auxiliary variable. However, it is still intractable to solve for $\mathbf{w}_k$ and $\mathbf{\Psi}$ directly since the last term of (\ref{f1}) is a sum of multiple fractions. Therefore, inspired by \cite{FP}, by introducing an auxiliary variable $\boldsymbol{\mu}\triangleq[\mu_{1}, \mu_2, \cdots, \mu_{K}]^{T}$, we apply a quadratic transform to the fractional terms to decouple the numerator and denominator, and further reformulate (\ref{f1}) as
\begin{align}
\label{f2}
   & f_{2}\left(\mathbf{w}_{k}, \boldsymbol{\Phi}, \mathbf{A}, \boldsymbol{\rho}, \boldsymbol{\mu}\right) \nonumber \\
    & =\sum_{k=1}^{K}\Big[\log _{2}\left(1+\mu_{k}\right)-\mu_{k}+2 \sqrt{1+\mu_{k}} \Re\left\{\eta_{k}^{*} \mathbf{h}_{k}^{H} \mathbf{w}_{k}\right\} \nonumber \\
&-\left|\eta_{k}\right|^{2}\big(\sum_{i=1}^{K}\left| \overline{\mathbf{h}}_{k}^{H}
 \mathbf{w}_{i}\right|^{2}+\left\|\mathbf{h}_{\mathrm{r}, k}^{H} \boldsymbol{\Psi}\right\|^{2} \sigma_{z}^{2}+\sigma_{k}^{2}\big)\Big].
\end{align}
Thus, the original problem (\ref{ori}) is reformulated as
\begin{subequations}
\label{final problem}
\begin{align}
    \max _{\mathbf{w}_{k}, \boldsymbol{\Phi}, \mathbf{A}, \boldsymbol{\rho}, \boldsymbol{\mu}} &~~ f_{2}\left(\mathbf{w}_{k}, \boldsymbol{\Phi}, \mathbf{A}, \boldsymbol{\rho}, \boldsymbol{\mu}\right) \\
    \text { s.t. } &~~(\text{\ref{ori b}})-(\text{\ref{ori d}}).
\end{align}
\end{subequations}
To effectively deal with the problem (\ref{final problem}), we propose applying the BCD method to alternately update $\boldsymbol{\rho}, \boldsymbol{\mu}, \mathbf{w}_{k}, \mathbf{A}$ and $\boldsymbol{\Phi}$ with fixed others, as demonstrated following.

\subsubsection{Update Auxiliary Variable $\boldsymbol{\rho}$}
Fixing $\mathbf{w}_{k}, \mathbf{A}$ and $\boldsymbol{\Phi}$, the optimal $\boldsymbol{\rho}$ is obtained by solving ${\partial f_{1}}/{\partial \rho_{k}}=0$, as given by
\begin{equation}
\label{rho}
\rho_{k}^{\star}=\frac{\big| \overline{\mathbf{h}}_{k}^{H}
 \mathbf{w}_{k}\big|^{2}}{\sum_{i=1, i \neq k}^{K}\big| \overline{\mathbf{h}}_{k}^{H}
 \mathbf{w}_{i}\big|^{2}+\big\|\mathbf{h}_{\mathrm{r}, k}^{H} \mathbf{\Psi}\big\|^{2} \sigma_{v}^{2}+\sigma^{2}}, ~~\forall k .
\end{equation}

\subsubsection{Update Auxiliary Variable $\boldsymbol{\mu}$}
Fixing $\boldsymbol{\rho}, \mathbf{w}_{k}, \mathbf{A}$ and $\boldsymbol{\Phi}$, the optimal $\boldsymbol{\mu}$ is obtained by solving ${\partial f_{2}}/{\partial \mu_{k}}=0$, as given by
\begin{equation}
\label{mu}
\mu_{k}^{\star}=\frac{\sqrt{1+\rho_{k}}  \overline{\mathbf{h}}_{k}^{H}
 \mathbf{w}_{k}}{\sum_{i=1}^{K}\big| \overline{\mathbf{h}}_{k}^{H}
 \mathbf{w}_{i}\big|^{2}+\big\|\mathbf{h}_{\mathrm{r}, k}^{H} \mathbf{\Psi}\big\|^{2} \sigma_{v}^{2}+\sigma^{2}}, ~~\forall k.
\end{equation}

\subsubsection{Update the BS beamforming vector $\mathbf{w}_{k}$}
For fixed $\boldsymbol{\rho}, \boldsymbol{\mu}, \mathbf{A}$ and $\boldsymbol{\Phi}$, by removing the constant term in (\ref{f2}) that is independent of $\mathbf{w}_{k}$ and defining $\mathbf{w} \triangleq[\mathbf{w}_{1}^{T}, \mathbf{w}_{2}^{T}, \cdots, \mathbf{w}_{K}^{T}]^{T}$, the optimization sub-problem with regard to $\mathbf{w}_{k}$ can be simplified as
\begin{subequations}
\label{updatewk}
\begin{align}
\max_{\mathbf{w}} & \enspace \Re\left\{\boldsymbol{\lambda}^{H} \mathbf{w}\right\}-\mathbf{w}^{H} \mathbf{E} \mathbf{w} \\
\text { s.t. } & \enspace \mathbf{w}^{H} \mathbf{w} \leq P_{\mathrm{BS}}, \\
& \enspace \left|\psi_{l}\right|^{2} \left(\mathbf{w}^{H} \mathbf{G}_{l} \mathbf{w}+\sigma_{v}^{2}\right) \leq P_{l},~~ \forall l, \\
& \enspace P_{\mathrm{in\_m}} \leq \mathbf{w}^{H} \mathbf{G}_{l} \mathbf{w} \leq P_{\mathrm{in\_max}}, ~~ \forall l,\label{inpowercons}
\end{align}
\end{subequations}
where for conciseness we define
\begin{subequations}\begin{align}
\boldsymbol{\lambda}&\triangleq\left[\boldsymbol{\lambda}_{1}^{T}, \boldsymbol{\lambda}_{2}^{T}, \cdots, \boldsymbol{\lambda}_{K}^{T}\right]^{T}, \\
\boldsymbol{\lambda}_{k}& \triangleq 2 \sqrt{1+\rho_{k}} \mu_{k} \overline{\mathbf{h}}_{k}, \\
\mathbf{E}&\triangleq\mathbf{I}_{K} \otimes\sum_{k=1}^{K}\left|\mu_{k}\right|^{2} \overline{\mathbf{h}}_{k} \overline{\mathbf{h}}_{k}^{H}, \\
\mathbf{G}_{l}&\triangleq\mathbf{I}_{K} \otimes\mathbf{h}_{\mathrm{d}, \mathrm{r}_{l}} \mathbf{h}_{\mathrm{d}, \mathrm{r}_{l}}^{H} .
\end{align}\end{subequations}
However, it is noted that the left part of the constraint (\ref{inpowercons}) is non-convex, making the problem (\ref{updatewk}) difficult to solve straightforwardly. As a consequence, by means of the MM algorithm based on the first-order Taylor expansion, we approximate the constraint as
\begin{equation}
	\mathbf{w}^{H} \mathbf{G}_{l} \mathbf{w} \geq \mathbf{w}^{H} \mathbf{G}_{l} \mathbf{w}+2\Re\left\{\mathbf{w}_{t}^{H} \mathbf{G}_{l} (\mathbf{w}-\mathbf{w}_{t})\right\},
\end{equation}
where $\mathbf{w}_{t}$ indicates the beamforming vector at the $t$-th iteration. Thus, problem (\ref{updatewk}) is reformulated as
\begin{subequations}
	\label{updatewknew}
	\begin{align}
		\max_{\mathbf{w}} & \enspace \Re\left\{\boldsymbol{\lambda}^{H} \mathbf{w}\right\}-\mathbf{w}^{H} \mathbf{E} \mathbf{w} \\
		\text {s.t.} & \enspace \mathbf{w}^{H} \mathbf{w} \leq P_{\mathrm{BS}}, \\
		& \enspace \left|\psi_{l}\right|^{2} \left(\mathbf{w}^{H} \mathbf{G}_{l} \mathbf{w}+\sigma_{v}^{2}\right) \leq P_{l},\quad \forall l, \\
		 &~~  \mathbf{w}^{H} \mathbf{G}_{l} \mathbf{w} \leq P_{\mathrm{in\_max}}, ~~ \forall l, \\
		 &  ~~\mathbf{w}^{H} \mathbf{G}_{l} \mathbf{w}+2\Re\left\{\mathbf{w}_{t}^{H} \mathbf{G}_{l} (\mathbf{w}-\mathbf{w}_{t})\right\} \geq P_{\mathrm{in\_m}}, ~~ \forall l.
	\end{align}
\end{subequations}
Obviously, problem (\ref{updatewknew}) is a typical quadratic constraint quadratic programming (QCQP) problem, and the optimal $\mathbf{w}$ can be easily obtained by existing toolboxes like CVX.
\subsubsection{Update RIS Reflection Amplification Matrix $\mathbf{A}$}
Fixing $\boldsymbol{\rho}, \boldsymbol{\mu}, \mathbf{w}_{k}$ and $\boldsymbol{\Phi}$, the reflection gain is related to the incident signal power as discussed in Section \ref{Amplify}. According to the simulation results in \cite{long}, we curve-fit the reflection gain of the $l$-th reflection amplifier in the linear amplification characteristic to obtain the expression as follows
\begin{equation}
\label{A}
\mathbf{A}_l = -0.195 p_{\mathrm{in},l} +22.46.
\end{equation}

\subsubsection{Update RIS Reflection Phase-Shift Coefficient $\psi$}
For fixed $\boldsymbol{\rho}, \boldsymbol{\mu}, \mathbf{w}_{k}$ and $\mathbf{A}$, as mentioned in Subsection $\mathrm{\uppercase\expandafter{\romannumeral2}}$-B, the equivalent channel $\overline{\mathbf{h}}_{k}^{H}$ can be re-expressed as
\begin{equation}
\overline{\mathbf{h}}_{k}^{H}=\boldsymbol{\psi}^{H} \operatorname{diag}\left(\mathbf{h}_{\mathrm{r}, k}^{H}\right) \mathbf{h}_{\mathrm{d,r}}.
\end{equation}
By removing the constant term in (\ref{f2}) that is independent of $\boldsymbol{\Psi}$, the optimization sub-problem with respect to $\boldsymbol{\psi}$ can be re-formulated as
\begin{subequations}
\label{updatepsi}
\begin{align}
\max _{\boldsymbol{\psi}} & \enspace \mathfrak{R}\left\{2 \boldsymbol{\psi}^{H} \boldsymbol{\Lambda}\right\}-\boldsymbol{\psi}^{H} \boldsymbol{\Pi} \boldsymbol{\psi} \\
\text { s.t. } & \enspace \left|\boldsymbol{\psi}_{l}\right|^{2} \leq \frac{P_{l}}{\sum_{k=1}^{K}|\mathbf{h}_{\mathrm{d,r}(l)}^{H} \mathbf{w}_{k}|^{2}+\sigma_{v}^{2}}, \quad \forall l,
\end{align}
\end{subequations}
where for brevity we define
\begin{subequations}\begin{align}
\label{Lambda}
\boldsymbol{\Lambda}& \triangleq \sum_{k=1}^{K} \sqrt{\left(1+\rho_{k}\right)} \operatorname{diag}\left(\mu_{k} \mathbf{h}_{\mathrm{r}, k}^{H}\right) \mathbf{h}_{\mathrm{d,r}} \mathbf{w}_{k},\\
\label{Pi}
\boldsymbol{\Pi} &\triangleq \sum_{k=1}^{K}\left|\mu_{k}\right|^{2}\sum_{i=1}^{K} \operatorname{diag}\left(\mathbf{h}_{\mathrm{r}, k}^{H}\right) \mathbf{h}_{\mathrm{d,r}} \mathbf{w}_{i} \mathbf{w}_{i}^{H} \mathbf{h}_{\mathrm{d,r}}^{H} \operatorname{diag}\left(\mathbf{h}_{\mathrm{r}, k}\right) \nonumber \\
& + \sum_{k=1}^{K}\left|\mu_{k}\right|^{2} \operatorname{diag}\left(\mathbf{h}_{\mathrm{r}, k}^{H}\right) \operatorname{diag}\left(\mathbf{h}_{\mathrm{r}, k}\right) \sigma_{v}^{2}.
\end{align}
\end{subequations}
We also note that problem (\ref{updatepsi}) is a QCQP problem whose optimal solution can be obtained by CVX.
\begin{algorithm}[!t]
    \caption{Pratical Joint BS Beamforming and RIS Reflection Precoding Design.}
    \label{alg}
    \begin{algorithmic}[1]
    \begin{small}
    \REQUIRE $\mathbf{h}_{\mathrm{d}, k}, \mathbf{h}_{\mathrm{r}, k}, \mathbf{h}_{\mathrm{d,r}}, P_{\mathrm{BS}}, P_{l}, \sigma_{v}^{2}, \sigma^{2}$.
    \ENSURE $\mathbf{w}_{k}^{\star}, \mathbf{A}^{\star}, \mathbf{\Phi}^{\star}, \forall k$.
        \STATE {Initialize $\mathbf{w}_{k}, \mathbf{A}, \boldsymbol{\Phi}, \forall l$. }
        \WHILE{no convergence of sum-rate}
            \STATE Update $\mathbf{\rho}$ by (\ref{rho});
            \STATE Update $\mathbf{\mu}$ by (\ref{mu});
            \STATE Update $\mathbf{w}_{k}$ by (\ref{updatewk});
            \STATE Update $\mathbf{A}$ by (\ref{A});
            \STATE Update $\boldsymbol{\psi}$ by (\ref{updatewk});
            \STATE Calculate sum-rate by (\ref{ori a}).
        \ENDWHILE
        \STATE Return $\mathbf{w}_{k}^{\star} = \mathbf{w}_{k}, \mathbf{A}^{\star} = \mathbf{A}$, and obtain $\mathbf{\Phi}^{\star}$ from $\boldsymbol{\psi}^{\star} = \boldsymbol{\psi}$.
        \end{small}\vspace{-0.0 cm}
    \end{algorithmic}
\end{algorithm}

Based on the above derivations, the overall joint transmit precoding and active RIS reflection beamforming design algorithm is summarized in Algorithm \ref{alg}.
By appropriately initializing the auxiliary variables $\boldsymbol{\rho}$ and $\boldsymbol{\mu}$, the BS transmit beamforming $\mathbf{w}_{k}$, RIS reflection amplification matrix $\mathbf{A}$ and the reflection coefficient $\boldsymbol{\psi}$ are iteratively updated, until the sum-rate of the MU-MISO system converges. 

\section{Simulation Results}
In this section, we illustrate simulation results to demonstrate the effectiveness of the proposed practical-based joint beamforming design algorithm. For the simulation setup, we assume that the BS equipped with $M=4$ antennas is located at $\left(0m, -40m\right)$ to serve $K=4$ users with an active RIS having $L=64$ elements. The active RIS is located at $\left(400m, 15m\right)$ to assist the MU-MISO communication system and the users are set to be randomly distributed in a circle with a radius of 8 meters from the center $\left(400m, 0m\right)$. A prevalent path-loss model is employed to indicate the large-scale fading: $\mathrm{PL}(d) = C_{0}(1/d)^{\alpha}$, where $C_{0} = -30 \mathrm{dB}$ and $\alpha$ denotes the path loss exponent. In particular, the path loss exponents for the channel between the BS and RIS, as well as the channel between RIS and users, are set as 3.2 and 2.7, respectively. The Rician fading channel model, which consists of both LoS and non-LoS (NLoS) components, is utilized to characterize the BS-RIS channel, which can be expressed as
\begin{equation}
\mathbf{H}_{\mathrm{R}} = \sqrt{\frac{\beta}{\beta+1}} \mathbf{H}_{\mathrm{LoS}}+\sqrt{\frac{1}{\beta+1}} \mathbf{H}_{\mathrm{NLoS}},
\end{equation}
where $\beta$ is the Rician factor and set to 1. $\mathbf{H}_{\mathrm{LoS}}$ is the LoS exponent that can be derived from location information, and $\mathbf{H}_{\mathrm{NLoS}}$ denotes the Rayleigh fading component. In addition, we assume that the channel from the BS to the user yields to the Rayleigh fading model.
For simplicity, we assume that perfect channel state information (CSI) is acquired.
Furthermore, as the static power consumption of hardware circuits of RIS is neglected, the other parameters are set as: $\sigma_{v}^{2} = \sigma^{2} = -90\mathrm{dBm}$, $P_{\mathrm{BS}} = 10\mathrm{dBm}$ and $P_{l} = 0.1\mathrm{dBm}$.

\begin{figure}[!t]
  \vspace{-0.3cm}
  \centering
  \includegraphics[width=0.42\textwidth]{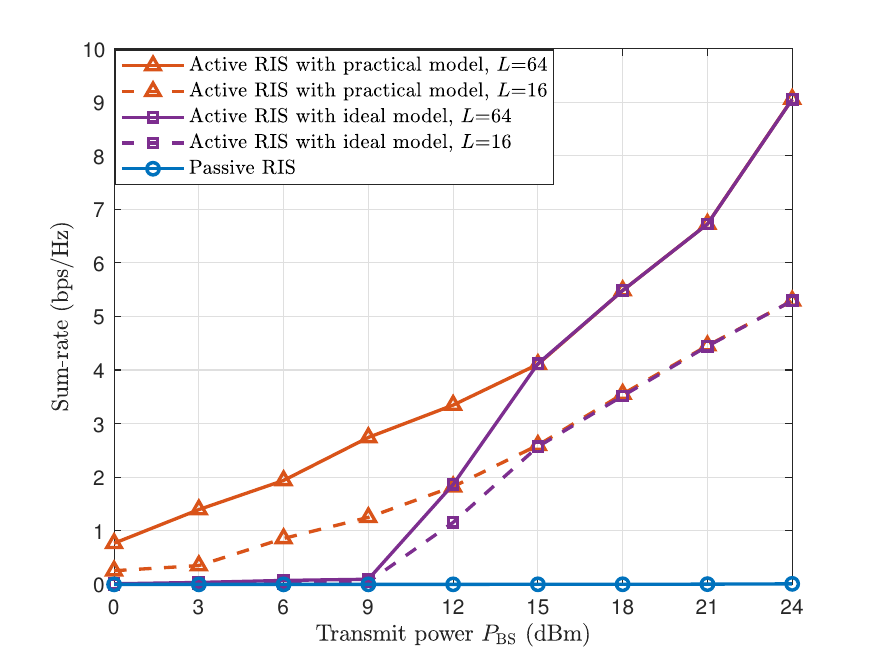}
  \caption{Sum-rate versus the transmit power budget $P_{\mathrm{BS}}$.}
  \label{fig:power}
  \vspace{-0.6cm}
\end{figure}

Fig. \ref{fig:power} illustrates the sum-rate performance versus the transmit power $P_{\mathrm{BS}}$ for \textit{i}) the active RIS-assisted system optimized using the proposed practical amplification model and designs; \textit{ii}) the active RIS-assisted system optimized using ideal amplification model; and \textit{iii}) passive RIS-assisted system. It is apparent that the higher transmit power enables the higher sum-rate of two active RIS-assisted systems, while the passive RIS-assisted system has almost zero performance improvement due to the double fading effect. Moreover, it is noteworthy that, with a transmit power $P_{\mathrm{BS}}$ less than 9dBm, the joint beamforming design using the ideal model exhibits a similar performance to the passive RIS. This is because that, with a limited transmit power budget, the design using the ideal model
can not provide strong enough incident signal power to trigger the amplification mechanism of the reflective amplifier. Consequently, the active RIS can only reflect the incident signal rather than amplify it, and thus  performs barely like a passive RIS. However, the active RIS can effectively operate within a low transmit power range through the utilization of our proposed model and algorithm.
As the transmit power lays in $\left[\mathrm{9dBm, 15dBm}\right]$, the power of the incident signal on the active RIS increases. Our proposed algorithm using the practical model still outperforms that using the ideal model because its incident signal power is still not strong enough to fully trigger the amplification mechanism.
When the transmit power is greater than 15dBm, the incident signal power is sufficient for the reflective amplifier to linearly amplify the incident signal. As a consequence, the designs with both ideal and practical model exhibits the same performance.

\begin{figure}[!t]
\vspace{-0.3cm}
\centering
  \includegraphics[width=0.42\textwidth]{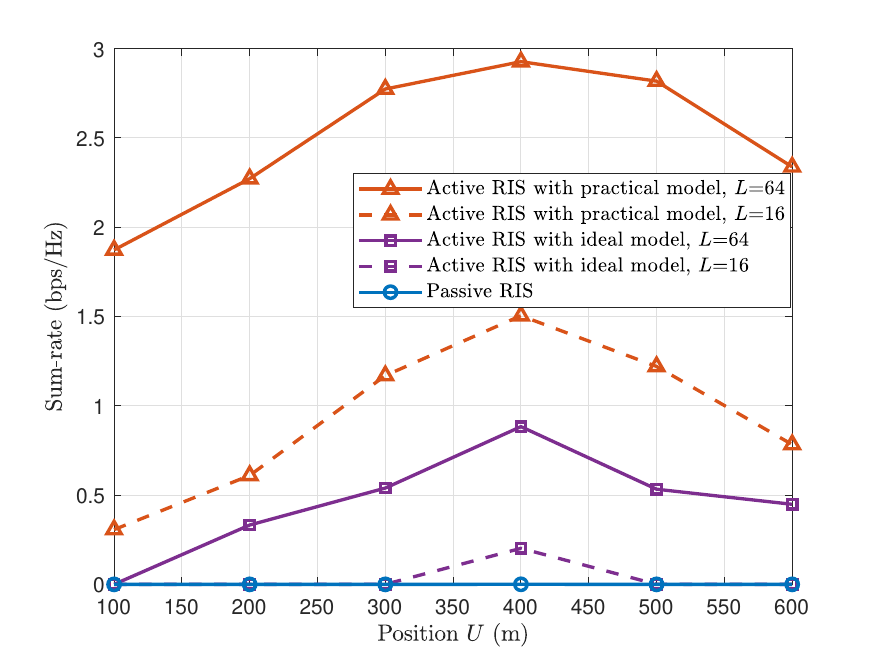}
  \caption{Sum-rate versus the position of users.}
  \label{fig:position}
\end{figure}

In Fig. \ref{fig:position}, we present the influence of the user's position on the sum-rate performance. It is found that the proposed practical model and design allow the active RIS-assisted system to exhibit better performance than the other two systems, regardless of the user's position. For $U = 400m$, the proposed beamforming designs achieve  231\% and 650\% gain compared to the active RIS designs using the ideal model at \textit{L}=64 and \textit{L}=16, respectively.
\begin{figure}[!t]
\vspace{-0.2cm}
\centering
  \includegraphics[width=0.42\textwidth]{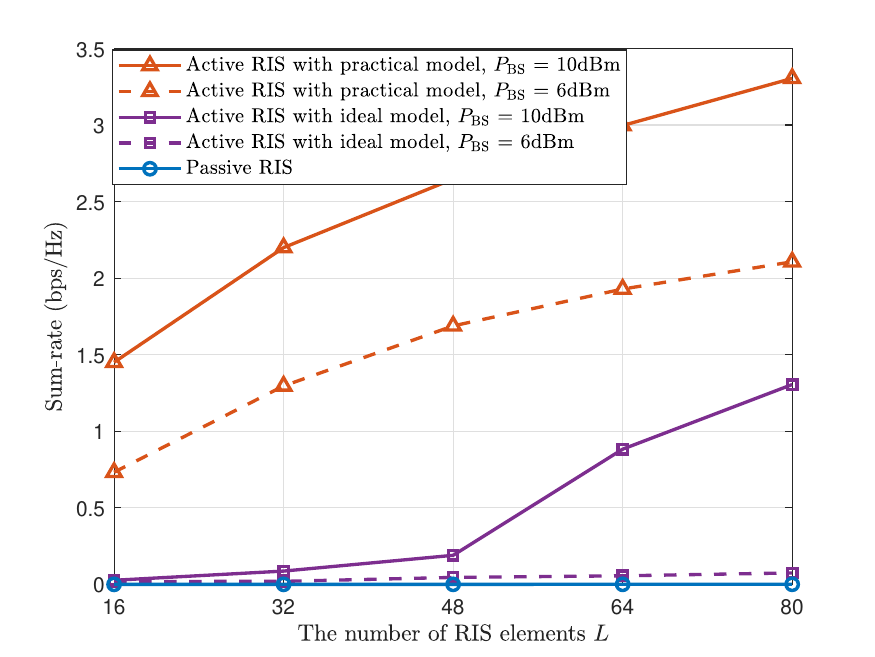}
  \caption{Sum-rate versus the number of RIS elements.}
  \label{fig:number}
  \vspace{-0.5cm}
\end{figure}

Finally, Fig. \ref{fig:number} illustrates the impact of the number of RIS elements on the performance. As the number of RIS elements $L$ increases, the proposed beamforming design using the practical model always outperforms the design using the ideal model. Particularly, the proposed design displays remarkable benefits when $P_{\mathrm{BS}}$ = 6dBm, while the ideal model-based design varies slightly with the number of elements due to the limited transmit power and incident signal power. It should be noted that with an increasing number of elements, the static power consumed by the hardware circuits of active RIS should not be negligible. Thus, a suitable number of elements will need to be chosen to maximize the advantages offered by this practical active RIS model.
Moreover, as shown in the above figures, passive RIS barely yields almost zero sum-rate, which indicates that it fails to serve communication systems due to the severe double fading effect under those scenarios.

\section{Conclusions}
In this paper, we presented a more practical amplification model based on the hardware characteristics of active RIS, and then proposed an efficient joint transmit beamforming and RIS reflection precoding design algorithm for the active RIS-assisted MU-MISO communication system.
The FP, MM, and BCD methods were utilized to tackle the resulting complicated optimization problem. Simulation results revealed that the design using the proposed practical amplification model provides a sum rate gain of up to 231\% in comparison with the counterpart based on the ideal model.

\end{document}